\documentclass[runningheads]{llncs}
\usepackage[T1]{fontenc}
\usepackage{ulem}
\usepackage{amsmath}
\usepackage{amssymb}
\usepackage{algorithm}
\usepackage{algorithmic}
\usepackage{multirow}
\usepackage{graphicx} 
\usepackage{booktabs} 
\usepackage{arydshln}
\usepackage{graphicx}
\usepackage{subcaption}
\usepackage{textcomp}
\newcommand{\hotflip}{HotFlip}

\newcommand{\myparagraph}[1]{\noindent \textbf{#1:}}

\newcommand{\best}[1]{\pmb{#1}}
\newcommand{\std}[1]{\scriptsize{(#1)}\normalsize}
\newcommand{\phantomstd}{\phantom{\std{00.0}}}
\newcommand{\nostd}{\scriptsize{(\hspace{0.382em}\textemdash\hspace{0.382em})}\normalsize}

\usepackage{hyperref}
\usepackage{color}
\hypersetup{
    colorlinks=true,
    urlcolor=blue, 
}

\begin{document}
\title{Reproducing HotFlip for Corpus Poisoning Attacks in Dense Retrieval}
%
%
\author{Yongkang Li\orcidID{0000-0001-6837-6184} \and
Panagiotis Eustratiadis\orcidID{0000-0002-9407-1293} \and
Evangelos Kanoulas\orcidID{0000-0002-8312-0694}}

%
%
\institute{University of Amsterdam, Amsterdam, The Netherlands \\
\email{\{y.li7,p.efstratiadis,e.kanoulas\}@uva.nl}}
%
\maketitle              
\begin{abstract}
\hotflip{} is a topical gradient-based word substitution method for attacking language models. Recently, this method has been further applied to attack retrieval systems by generating malicious passages that are injected into a corpus, i.e., corpus poisoning.
However, \hotflip{} is known to be computationally inefficient, with the majority of time being spent on gradient accumulation for each query-passage pair during the adversarial token generation phase, making it impossible to generate an adequate number of adversarial passages in a reasonable amount of time. 
Moreover, the attack method itself assumes access to a set of user queries, a strong assumption that does not correspond to how real-world adversarial attacks are usually performed.
In this paper, we first significantly boost the efficiency of \hotflip{}, reducing the adversarial generation process from 4 hours per document to only 15 minutes, using the same hardware.
We further contribute experiments and analysis on two additional tasks: (1) transfer-based black-box attacks, and (2) query-agnostic attacks.
Whenever possible, we provide comparisons between the original method and our improved version.
Our experiments demonstrate that \hotflip{} can effectively attack a variety of dense retrievers, with an observed trend that its attack performance diminishes against more advanced and recent methods.
Interestingly, we observe that while \hotflip{} performs poorly in a black-box setting, indicating limited capacity for generalization, in query-agnostic scenarios its performance is correlated to the volume of injected adversarial passages.

\end{abstract}
\section{Introduction}

We consider adversarial attacks on information retrieval (IR) systems that encode queries and passages as dense vectors, and use their cosine similarity or inner product to model query-passage relevance. We focus on dense retrieval methods that employ a transformer-based bi-encoder architecture~\cite{HofstatterLYLH21_tasb_dense_retrieval,izacard2021contriever,KarpukhinOMLWEC20_DPR,LinALOLMY023_dragon}, as they are the most commonly used; combining computational efficiency and high performance.
In this paper, we reproduce the work of Zhong et al.~\cite{ZhongHWC23_Poisoning}, who proposed an adversarial passage generation pipeline that uses gradient-based word substitution, for corpus poisoning. The word-substitution method they used is termed \hotflip{}~\cite{EbrahimiRLD18_hotflip}, and it is the essence of our reproducibility study.

\begin{figure}[t]
\centering
  \includegraphics[width=\columnwidth]{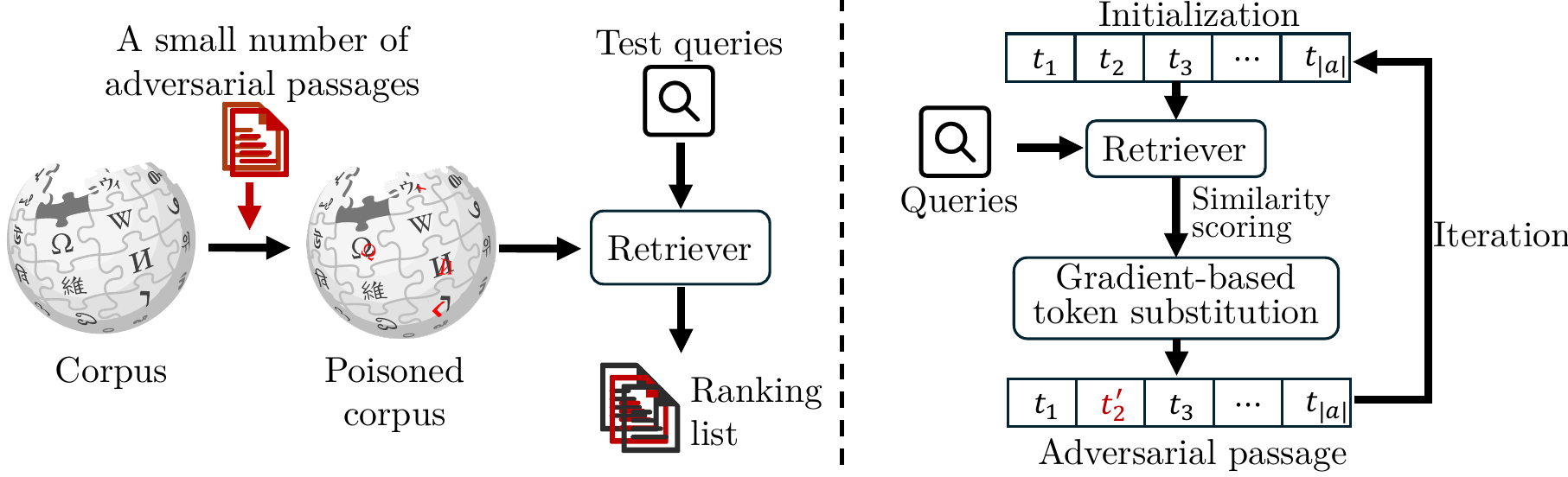}
  \caption{Illustration of corpus poisoning with \hotflip{}. \textbf{Left:} The corpus poisoning attack aims to inject a small number of adversarial passages into the corpus to maximize their impact on ranking. \textbf{Right:} A simplified pipeline for generating adversarial passages with \hotflip{}. In each iteration, the retrieval model calculates the similarity between an adversarial passage and some queries. It then selects a token at random and replaces it with a loss-maximizing token.}
  \label{fig:task illustration}
\end{figure}

Corpus poisoning attacks are some of the most dangerous ways to target IR systems~\cite{liu2024robust,DBLP:journals/corr/abs-2407-06992}, as only a small number of adversarial documents is enough to contaminate an entire corpus~(Figure~\ref{fig:task illustration}, left). It is fairly common to see instances of corpus poisoning in practice; we often see such attacks in recommendation engines when, for example, maliciously generated product descriptions are injected into a website's catalog to compromise the trustworthiness of an online vendor.

Zhong et al.\cite{ZhongHWC23_Poisoning} demonstrate that \hotflip{} can produce very strong adversarial passages that can substantially damage the performance of dense retrievers. Our first contribution in this paper, is to study the reproducibility of their results and confirm that they are largely reproducible.
However, to some extent, they trade off good attack performance for lack of realism. Our further contributions are the extensive analysis and experimentation with these trade-offs.

Specifically, \hotflip{} is known to be computationally inefficient~\cite{abs-2402-12784_Vec2Text_Understanding}, 
with the majority of time being spent on gradient accumulation for each query-passage pair during the adversarial token generation phase.
In our experiments, an average of 4 GPU/h are needed to generate a single 50-token-long adversarial passage, on an NVIDIA L40 GPU. This leads us to our first research question (\textbf{RQ1}): Can the efficiency of \hotflip{} be improved without sacrificing performance?
We optimize the step of gradient calculations for every query-passage pair by clustering the queries and considering the embedding of their centroid. 
Our extensive experiments show that our optimization strategy has achieved benefits in both efficiency and effectiveness.

Furthermore, the original paper only considers gradient-based, white-box attacks in their experiments, where the attacker has complete unconstrained access to the target model. However, as mentioned, corpus poisoning attacks are applicable in many cases where details about the target model cannot be known. Our second research question (\textbf{RQ2}) is: How does \hotflip{} perform in a black-box attack setting? We experiment with a transfer-based attack, and our results suggest that \hotflip{} cannot generalize well across models.

Finally, the adversarial passage generation pipeline of Zhong et al.~\cite{ZhongHWC23_Poisoning} relies on prior knowledge of queries for which the attack is optimized (Figure~\ref{fig:task illustration}, right). However, prior knowledge of user queries is not a realistic signal that we can rely on; especially in systems that value privacy. Our final research question (\textbf{RQ3}) is: How does \hotflip{} perform without prior query knowledge? We modify the method to use other passages instead of queries, and propose a new paradigm to measure the performance of \hotflip{}. 
We found that even corrupting as little as 0.01\% of the corpus in this way can sometimes decrease the effectiveness of the IR system, depending on the target model and the size of the corpus.

Overall, our work yields two important outcomes: (i) it leads to a better understanding of the \hotflip{}-based attack proposed by Zhong et al.~\cite{ZhongHWC23_Poisoning} under more realistic scenarios, and (ii) it enables fast generation of high-quality adversarial documents, that can then be used as negative examples (e.g.,~\cite{ParkC19_adv_training,pmlr-v119-zhang20b}) to train more robust dense retrievers.
\looseness=-1

\section{Related Work}

\myparagraph{\hotflip{}}
\hotflip{}~\cite{EbrahimiRLD18_hotflip} is a widely used token substitution method for generating adversarial examples to target language models in a white-box setting. It models text operations, i.e., ``flips'', as vectors, and corrupts text by applying the operations that maximize the error of the target model, assuming access to its gradients.
Subsequent research has applied this method to various scenarios, such as machine translation~\cite{EbrahimiLD18}, explainability~\cite{WallaceTWSGS19,WangTW020},  counterfactual instance generation~\cite{emnlp_FernP21,sigir_0006SA22}, and data augmentation~\cite{pmlr-v119-zhang20b}.

\myparagraph{Attacks in Dense Retrieval}
Dense retrieval models have become widely used in IR systems. However, transformer-based dense retrieval models, such as DPR~\cite{KarpukhinOMLWEC20_DPR}, Contriever~\cite{izacard2021contriever}, TAS-B~\cite{HofstatterLYLH21_tasb_dense_retrieval}, and Dragon$+$~\cite{LinALOLMY023_dragon}, are inherently vulnerable to adversarial attacks~\cite{ChenHY0S23,10.1145/3548606.3560683_Order-Disorder,10.1145/3583780.3614793_MCARA,Liu0GR0FC23_TARA,Liu0GRFC24,abs-2008-02197_one_word,song2020adversarial,su2024corpus,10.1145/3576923_PRADA} due to the nature of neural networks~\cite{GoodfellowSS14,Szegedy13Intriguing}, which can severely damage their retrieval performance~\cite{10.1145/3583780.3614793_MCARA,10.1145/3576923_PRADA}.
Most attacks focus on word substitution, employing a gradient-guided greedy substitution strategy similar to \hotflip{}. For instance, PRADA~\cite{10.1145/3576923_PRADA} identifies key tokens by analyzing gradient magnitudes and uses the slope of the loss function to find synonyms for substitution. Similarly, MCARA~\cite{10.1145/3583780.3614793_MCARA} utilizes a view-wise contrastive loss to compute gradients, and applies a word selection and synonym substitution method similar to PRADA. The latest of these models, RL-MARA~\cite{Liu0GRFC24}, employs a policy network coupled with reinforcement learning to identify vulnerabilities in the target document, guiding subsequent attack steps.
The method studied in this paper by Zhong et al.\cite{ZhongHWC23_Poisoning}, introduces a novel pipeline that focuses on contaminating the corpus with a small number of harmful passages, i.e., is a corpus poisoning method. A similar method, AGGD~\cite{su2024corpus}, is also worth mentioning; it adopts a different strategy for selecting token candidates, but follows the same overall approach.

\section{Method Overview}

\begin{algorithm}[t]
\caption{Corpus Poisoning Attack with \hotflip{}~\cite{ZhongHWC23_Poisoning}} %
\label{alg:\hotflip{}} %
\begin{algorithmic}[1] 
\REQUIRE Query set~\(Q\), corpus~\(\mathcal{C}\), similarity function~\(\text{sim}(\cdot)\), max iterations \(I_{max}\) %
\ENSURE Adversarial passage \(a\) %
\STATE Randomly initialize~\(a\)~from~\(\mathcal{C}\): \(a\leftarrow[t_1,t_2,\dots,t_{|a|}]\)   %
\FOR{$I_{\text{max}}$ {steps}}
\STATE Select a random token \(t_i \in a\), calculate Equation~\ref{eq:eq2} and get top-\textit{n} candidates.
\FOR{each  $t_i'$ in the top-\textit{n} candidates}
    \STATE Replace $t_i\leftarrow t_i'$, we get  \(a'\leftarrow [t_1,t_2,\dots,t_i',\dots,t_{|a|}] \)
\ENDFOR
\STATE Calculate Equation~\ref{eq:eq1} to select best $t_i'$  and its  corresponding~\(a'\)
\IF{\( \sum_{q \in Q} \text{sim}(q, a') > \sum_{q \in Q} \text{sim}(q, a)\)}
\STATE \(a\leftarrow a'\)
\ENDIF
\ENDFOR
\RETURN \(a\)%
\end{algorithmic}
\end{algorithm}

In this section, we begin by defining the task and outlining the objectives. Then, we discuss the adversarial passage generation pipeline proposed by Zhong et al.~\cite{ZhongHWC23_Poisoning}. Lastly, we present our optimization strategy, which uses the embedding centroids of clustered queries to improve efficiency.

\subsection{Corpus Poisoning Attack}
For a dual-encoder dense retrieval system, the query encoder \(E_q(\cdot)\) encodes a user query \(q\) as the embedding \(E_q(q)\in \mathbb{R}^{d}\), while the passage encoder \(E_p(\cdot)\)  will encode the corpus \(\mathcal{C}=\left\{p_1,p_2,\dots,p_{|\mathcal{C}|}\right\}\). The retriever aims to return top-\(k\) most similar passages based on their similarity to the query, typically calculated as the dot product of their embeddings: \(\text{sim}(q,p) = E_q(q)^\top E_p(p) \). 

The goal of corpus poisoning is to compromise the corpus and mislead the retriever by generating and injecting a small set of adversarial passages \(\mathcal{A}=\left\{a_1, a_2,\dots, a_{|\mathcal{A}|}\right\}\) into the corpus, where \(|\mathcal{A}|\ll|\mathcal{C}|\).
Each adversarial passage $a$ is composed of a sequence of tokens $a =[t_1, t_2,\cdots,t_{|a|} ] $. 
These adversarial passages are designed to be highly similar to specific queries, ensuring they are returned by the retriever. 
To generate adversarial passages in \(\mathcal{A}\), Zhong et al.\cite{ZhongHWC23_Poisoning}  assume access to a complete training set of queries \(\mathcal{Q}\), and expect \(\mathcal{A}\) will also be effective for unseen test queries.

\subsection{Adversarial Passage Generation Pipeline}\label{sec:Query-based Optimization}

Following the approach described by Zhong et al.~\cite{ZhongHWC23_Poisoning},  to generate $k$ adversarial passages to attack the corpus, we first apply $k$-means clustering on all training queries, obtaining multiple clusters. For each cluster and its query set $Q = \{q_1, q_2, \cdots,q_{|Q|}\}$, we generate one adversarial passage $a$ that is ranked highly by the retriever in the ranking results.
To mislead the model, we maximize the similarity between $a$ and the queries in \(Q\), as shown in Equation~\ref{eq:eq1}.
\begin{align}
    a &= \arg \max_{a'} \frac{1}{|Q|} \sum_{q \in Q}  \text{sim}(q,a') \label{eq:eq1} 
\end{align}

To generate each adversarial passage, Zhong et al.~\cite{ZhongHWC23_Poisoning}  initialize $a$ using a random passage from the corpus. A token \(t_i\in a\) is then selected, and the gradient \(\nabla_{e_{t_i}} \text{sim}(q, a) \) is computed with respect to its token embedding \(e_{t_i}\).
Using \hotflip{}, the approximation of the retriever output when replacing \(t_i\) with a better token \(t_i'\) is \(e_{t_i'}^\top \nabla_{e_{t_i}} \text{sim}(q, a)\). By maximizing Equation~\ref{eq:eq1}  to increase similarity to \(Q\),  we  optimize Equation~\ref{eq:eq2} as follows:
\begin{align}
            a' = \arg \max_{t_i' \in \mathcal{V}} \frac{1}{|Q|} \sum_{q \in Q} e_{t_i'}^\top \nabla_{e_{t_i}} \text{sim}(q, a) \label{eq:eq2} 
\end{align}
where \(\mathcal{V}\) is the vocabulary.
Each iteration repeats this process: first, we use Equation~\ref{eq:eq2} to identify the top-\( n \) candidate replacement tokens \( t_i' \). Since Equation~\ref{eq:eq2} only approximates the retriever's output, we then replace \( t_i \) with each candidate \( t_i' \) to generate \( a' \), selecting the best \( a' \) that satisfies Equation~\ref{eq:eq1}.   The entire process is outlined in Algorithm~\ref{alg:\hotflip{}}.

\subsection{Pipeline Optimization Strategy}\label{sec:Pipeline Optimization Strategy}

However,  calculating the similarity between $a$ and all queries \(q \in Q\) is impractical when \(|Q|\) is large.  To simplify this, Zhong et al.~\cite{ZhongHWC23_Poisoning} used a sampled batch of queries \(Q_b \in Q\). Despite limiting the calculations to this batch, the algorithm remains extremely time-consuming. 
To improve efficiency, we use the mean embedding of all queries within each batch of the cluster as the attack target. As a result, Equation~\ref{eq:eq1} is modified to Equation~\ref{eq:eq3}, and Equation~\ref{eq:eq2} is modified to Equation~\ref{eq:eq4}, as shown below:
\begin{align}
    a    &\simeq  \arg \max_{a'} \bar{E}_q(Q_b)^\top E_p(a')\label{eq:eq3} \\
    a' &\simeq \arg \max_{t_i' \in \mathcal{V}} e_{t_i'}^\top \nabla_{e_{t_i}}{ \left(\bar{E}_q(Q_b)^\top E_p(a)  \right)} \label{eq:eq4} 
\end{align}
where $\bar{E}_q(Q_b) \in \mathbb{R}^{d}$ is the mean embedding of all queries in $Q_b$. Since  \(\bar{E}_q(Q_b)\)  can be precomputed and stored during the data processing, there is no need to recalculate it during token searching, significantly reducing the overall complexity of the algorithm.
Our experiment results in Section~\ref{result:RQ1} show that combining this optimization strategy with batch processing reduced the runtime from 4 hours to 15 minutes. Moreover, the optimized pipeline maintains performance while significantly enhancing efficiency.

\section{Experiments Setup}

In this section, we outline the experimental setup, detailing the datasets used and our attack target models, which include a series of dense retrievers.
 We also describe our evaluation metrics and provide implementation details.

\subsection{Datasets}

Following Zhong et al.~\cite{ZhongHWC23_Poisoning}, we use two well-known datasets,  Natural Questions (\textbf{NQ}~\cite{KwiatkowskiPRCP19_NQ}) and  \textbf{MS MARCO}~\cite{bajaj2016ms} to reproduce the main results in Zhong et al. 
Additionally, for attacks without queries (RQ3), we  select  two datasets from the {BEIR}~\cite{thakur2021beir} benchmark: \textbf{ArguAna}~\cite{WachsmuthSS18_arguana} and \textbf{FiQA}~\cite{MaiaHFDMZB18_fiqa}. These two datasets were chosen for their smaller corpus size, which allows for generating a specific proportion of adversarial passages and enables a fair comparison of attack effectiveness, as detailed in Section~\ref{sec:RQ3_exp}. The dataset statistics are provided in Table~\ref{tab:datasets_details}.\looseness=-1

\begin{table*}[tbp] 
\caption{Statistics of datasets used in our work~\cite{thakur2021beir}. 
Avg. D/Q indicates the average number of relevant documents per query.} 
\centering
\renewcommand\arraystretch{1.3} %
\setlength{\tabcolsep}{0.5mm}{ %
\resizebox{1.\linewidth}{!}{
\begin{tabular}{lcccccccccccccc}
\toprule
 \multirow{2}{*}{Datasets}  & \multirow{2}{*}{Task} & \multirow{2}{*}{Domain} & \multirow{2}{*}{Title}    & \multirow{2}{*}{\#Corpus}    &\multicolumn{1}{c}{Train}  &\multicolumn{2}{c}{Test} &\multicolumn{2}{c}{Avg. Word Lengths} 
    \\\cmidrule(lr){6-6}\cmidrule(lr){7-8} \cmidrule(lr){9-10}
           &  &  &   & &\#Pairs  &\#Query &Avg.D/Q   &Query  &Document \\
\hline
NQ   &Question Answering  & Wikipedia & \checkmark  &2,681,468 &132,803 &3,452 &1.2 &9.16 &78.88   \\
MS MARCO   &Passage-Retrieval & Misc.  & \texttimes  &8,841,823 &532,761 &6,980  &1.1 &5.96&55.98 \\
\hdashline
ArguAna & Argument &Misc.  & \checkmark  &8,674 &--- &1,406 &1.0 &192.98 &166.80 \\
FiQA & Question Answering &Finance  & \texttimes  &57,638 &14,166 &648 &2.6 &10.77 &132.32 \\
\bottomrule
\end{tabular}}
}
 \label{tab:datasets_details}
\end{table*}

\subsection{Dense Retrievers And Evaluation Metrics}

\myparagraph{Dense Retrieval} Following  Zhong et al.~\cite{ZhongHWC23_Poisoning}, we use five popular bi-encoder models as retrievers: Contriever~\cite{izacard2021contriever} (pre-trained), Contriever-ms (fine-tuned on MS MARCO), DPR-nq~\cite{KarpukhinOMLWEC20_DPR} (trained on NQ),  DPR-mul (trained on multiple datasets), and ANCE~\cite{XiongXLTLBAO21_ance_raw}.
Additionally, we include two commonly used bi-encoder models for comparison: TAS-B~\cite{HofstatterLYLH21_tasb_dense_retrieval} (trained via knowledge distillation), and Dragon$+$\cite{LinALOLMY023_dragon}, a state-of-the-art (SOTA) bi-encoder model.
All retrieval models are sourced from public repositories without any modifications.

\myparagraph{Evaluation Metrics} Following  Zhong et al.~\cite{ZhongHWC23_Poisoning}, we evaluate the top-$k$ attack success rates on test queries. This is defined as the percentage of queries for which at least one adversarial passage appears in the top-$k$ retrieval results.  
For simplicity, we omit the percentage symbol \%  throughout all the result tables.

\subsection{Implementation Details}

Following Zhong et al.~\cite{ZhongHWC23_Poisoning}, all adversarial passages are initialized with a length of 50 tokens, i.e.,~\(|a|=50\). And we run a maximum of 5000 iterations for each generation,  i.e.,~\(I_{max}=5000\). We select the top-100~(\textit{n}=100) tokens as potential replacements, meaning that step 3 of Algorithm~\ref{alg:\hotflip{}} considers these top-100 candidates. 
It is important to note that in the implementation of  Zhong et al., typically only one batch of queries~\(Q_b\) is used per iteration, and the batch varies across iterations.
We use the following five random seeds for this work: 1999, 5, 27, 2016, and 2024.
The main experiments are implemented using Pytorch 2.1 on a Ubuntu server with NVIDIA L40 GPU ×8, AMD EPYC 9534 CPU ×4, and 760G memory. 
We also use the Dutch national supercomputer Snellius to accelerate the experiments. 
Our code is available at \url{https://github.com/liyongkang123/hotflip_corpus_poisoning}.

\section{Experimental Results}
In this section, we design experiments to address the three research questions of our reproducibility study and obtain the corresponding conclusions.

\begin{table*}[t]
\caption{
Top-20 attack success rates for in-domain experiments with all three implementations of \hotflip{}. We run the experiments with 5 random seeds and report the mean and standard deviation in parentheses. In the case of $|\mathcal{A}| = 50$, it is computationally unfeasible to the original \hotflip{} for all random seeds, thus \nostd{} indicates that these experiments were only run once.
}
\centering
\renewcommand\arraystretch{1.1}
\setlength{\tabcolsep}{1mm}{
\resizebox{1.\linewidth}{!}{
\begin{tabular}{llcccccc}
    \toprule
    \multirow{2}{*}{Retrievers} & \multirow{2}{*}{\parbox{1cm}{\centering \hotflip{}\\[0.3em]impl.}} & \multicolumn{3}{c}{$|\mathcal{A}|$ of NQ} & \multicolumn{3}{c}{ $|\mathcal{A}|$ of MS MARCO}  \\
    \cmidrule(lr){3-5} \cmidrule(lr){6-8}
    & & 1\(\,\uparrow\) & 10\(\,\uparrow\) & 50\(\,\uparrow\) & 1\(\,\uparrow\) & 10\(\,\uparrow\) & 50\(\,\uparrow\)  \\ 
    \midrule
    \multirow{3}{*}{Contriever}
    & Original & {\best{84.2}~\phantomstd{}} & {\best{98.1}~\phantomstd{}} & {\best{99.4}~\phantomstd{}} & {\best{75.2}~\phantomstd{}}  & {\best{92.2}~\phantomstd{}} & {\best{98.6}~\phantomstd{}} \\
    & Reproduced & 80.1~\std{\phantom{0}2.7} & 97.2~\std{\phantom{0}0.5} & 99.2~\nostd{} & 62.3~\std{\phantom{0}5.0} & 92.1~\std{\phantom{0}3.8}  & 97.7~\nostd{} \\
    & Ours & 73.1~\std{19.9} & 96.9~\std{\phantom{0}0.4} & {99.4}~\std{\phantom{0}0.2} & 56.3~\std{13.4} & 90.2~\std{\phantom{0}4.5} & 97.2~\std{\phantom{0}2.7}\\
    \hdashline
    \multirow{3}{*}{Contriever-ms}
    & Original & \phantom{0}0.5~\phantomstd{} & 52.5~\phantomstd{} & 80.9~\phantomstd{} & \phantom{0}2.4~\phantomstd{} & 20.9~\phantomstd{} & 34.9~\phantomstd{} \\
    & Reproduced & \best{48.4}~\std{\phantom{0}1.3} & \best{84.0}~\std{\phantom{0}1.6} & \best{96.0}~\nostd{} & \phantom{0}\best{9.8}~\std{\phantom{0}2.7} & 60.9~\std{\phantom{0}4.0} & \best{95.3}~\nostd{} \\
    & Ours & 44.6~\std{21.2} & 82.5~\std{\phantom{0}2.3} & 95.4~\std{\phantom{0}2.4} & \phantom{0}8.8~\std{\phantom{0}5.8} & \best{63.2}~\std{\phantom{0}8.1} & {90.7}~\std{\phantom{0}2.9} \\
    \hdashline
    \multirow{3}{*}{DPR-nq}
    & Original & \phantom{0}\best{0.0}~\phantomstd{} & \phantom{0}\best{3.8}~\phantomstd{} & \best{18.8}~\phantomstd{} & \phantom{0}\best{0.1}~\phantomstd{} & \phantom{0}\best{2.6}~\phantomstd{} & \best{13.9}~\phantomstd{} \\
    & Reproduced & \phantom{0}0.0~\std{\phantom{0}0.0} & \phantom{0}1.3~\std{\phantom{0}0.3} & \phantom{0}7.0~\nostd{} & \phantom{0}0.0~\std{\phantom{0}0.0} & \phantom{0}0.0~\std{\phantom{0}0.0} & \phantom{0}4.7~\nostd{} \\
    & Ours & \phantom{0}0.0~\std{\phantom{0}0.0} & \phantom{0}2.9~\std{\phantom{0}0.9} & 12.0~\std{\phantom{0}0.4} & \phantom{0}0.0~\std{\phantom{0}0.0} & \phantom{0}0.5~\std{\phantom{0}0.9} & \phantom{0}4.7~\std{\phantom{0}2.1} \\
    \hdashline
    \multirow{3}{*}{DPR-mul}
    & Original & \phantom{0}0.0~\phantomstd{} & 10.6~\phantomstd{} & 28.3~\phantomstd{} & \phantom{0}0.0~\phantomstd{} & \phantom{0}4.7~\phantomstd{} & 16.3~\phantomstd{} \\
    & Reproduced & \phantom{0}0.0~\std{\phantom{0}0.0} & \phantom{0}7.0~\std{\phantom{0}0.7} & 24.4~\nostd{} & \phantom{0}0.5~\std{\phantom{0}1.0} & \phantom{0}2.3~\std{\phantom{0}2.5} & 25.6~\nostd{} \\
    & Ours & \phantom{0}\best{0.0}~\std{\phantom{0}0.0} & \best{11.3}~\std{\phantom{0}0.7} & \best{31.5}~\std{\phantom{0}0.7} & \phantom{0}\best{0.5}~\std{\phantom{0}0.9} & \phantom{0}\best{7.4}~\std{\phantom{0}0.9} & \best{30.2}~\std{\phantom{0}3.6} \\
    \hdashline
    \multirow{3}{*}{ANCE}
    & Original & \phantom{0}1.0~\phantomstd{} & 14.7~\phantomstd{} & 34.3~\phantomstd{} & \phantom{0}0.0~\phantomstd{} & \phantom{0}2.3~\phantomstd{} & 11.6~\phantomstd{} \\
    & Reproduced & \phantom{0}1.0~\std{\phantom{0}0.2} & 14.3~\std{\phantom{0}0.5} & 34.5~\nostd{} & \phantom{0}0.0~\std{\phantom{0}0.0} & \phantom{0}0.9~\std{\phantom{0}1.1} & 11.6~\nostd{}  \\
    
    & Ours & \phantom{0}\best{1.9}~\std{\phantom{0}0.1} &\best{22.2}~\std{\phantom{0}0.2} & \best{43.4}~\std{\phantom{0}0.7} & \phantom{0}\best{0.5}~\std{\phantom{0}0.9} & \phantom{0}\best{0.9}~\std{\phantom{0}1.1} & \best{19.5}~\std{\phantom{0}3.2} \\
    \bottomrule
\end{tabular}
}
}    
\label{tab:results_successrate}
\end{table*}

\subsection{RQ1: Can the efficiency of \hotflip{} be improved without sacrificing performance?}\label{result:RQ1}

We reproduce the main \hotflip{} attack results of Zhong et al.~\cite{ZhongHWC23_Poisoning} in two ways and perform a comparison, producing three types of results in total:\\
(1) \textbf{Original}: the original \hotflip{} results reported by Zhong et al.\\
(2) \textbf{Reproduced}: the results we reproduced by running the code of Zhong et al.\looseness=-1\\
(3) \textbf{Ours}:  the results from our optimized pipeline based on Equations~\ref{eq:eq3} and~\ref{eq:eq4}.

\subsubsection{Attack  Effectiveness Analysis}
We generate $|\mathcal{A}| \in \{1, 10, 50\}$ adversarial passages using the training query sets of NQ or MS MARCO and evaluate them on the corresponding held-out test queries.
The experimental results are shown in Table~\ref{tab:results_successrate}, from which we can observe that:

$\bullet$ 
The results of \textbf{Reproduced}, obtained by repeating the experiment five times with different random seeds, are generally consistent with those of the \textbf{Original}, indicating that the findings of Zhong et al. are reproducible. However, some minor discrepancies remain.
Specifically, the \textbf{Reproduced} results for  Contriever, DPR-nq, DPR-multi, and ANCE are lower than those reported in the \textbf{Original}. However, for Contriever-ms, \textbf{Reproduced} significantly outperforms the \textbf{Original}. 
Surprisingly, we find that \textbf{Reproduced} achieves a top-20 attack success rate of 48.4\% with $|\mathcal{A}|=1$ on the NQ dataset, while the original paper reported only 0.5\%.
We suspect a potential typo in the results reported by Zhong et al.~\cite{ZhongHWC23_Poisoning},  but have not yet received a response after contacting the authors via email for clarification.

$\bullet$ Comparing \textbf{Reproduced} and \textbf{Ours}, we observe that \textbf{Ours} slightly underperforms the original implementation only when attacking the Contriever and Contriever-ms. However, for the other three retrievers, DPR-nq, DPR-multi, and ANCE, \textbf{Ours} consistently achieves the best results. These findings indicate that the optimization strategy using average embeddings is highly effective.\looseness=-1

$\bullet$ Our optimized pipeline \textbf{Ours} shows high variance when $|\mathcal{A}|=1$,
which arises from the random initialization of adversarial passages.
Moreover, \textbf{Reproduced} computes the sum of gradients from a batch of queries, leading to greater stability during token searching, while \textbf{Ours} uses only a single gradient, resulting in higher variance.
However, we observe a significant reduction in variance as $|\mathcal{A}|$ increases, indicating that \textbf{Ours} achieves more stable performance when generating a larger number of adversarial samples.

\begin{figure}[t]
\centering
  \includegraphics[width=\columnwidth]{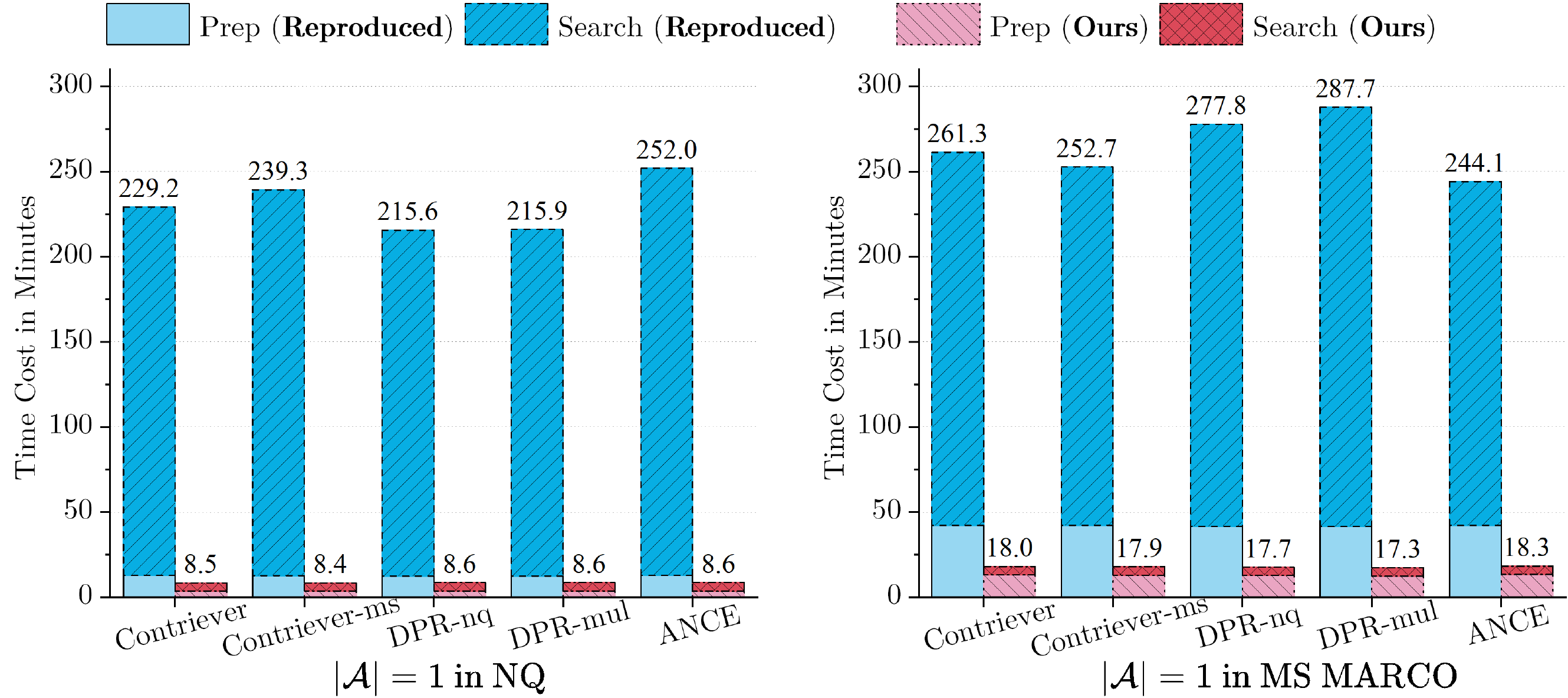}
  \caption{In-domain attack time cost~(in minutes) to generate a 50-tokens adversarial passage between two implementations of \hotflip{}: \textbf{Reproduced} and \textbf{Ours}.
  }
  \label{fig:results_time}
\end{figure}

\subsubsection{Attack  Efficiency Analysis}

To compare the efficiency between \textbf{Ours}, the pipeline optimized using Equations~\ref{eq:eq3} and~\ref{eq:eq4}, and \textbf{Reproduced}, the code from Zhong et al.\cite{ZhongHWC23_Poisoning}, we record the time cost for the aforementioned experiments.
We divide the time of generating an adversarial passage into two components: 

(1)\textbf{Prep}: The time cost for data loading, embedding queries, clustering, etc.

(2)\textbf{Search}: The time cost for iterative token substitution and searching.

The time cost results are presented in Figure~\ref{fig:results_time}, where we observe that:

$\bullet$ For the same process, the time required to generate an adversarial passage depends solely on the size of the dataset and the retrieval model used. 
In this experiment, all five retrievers are of BERT-base size; thus,  with a fixed dataset, the \textbf{Search} times for  \textbf{Reproduced} across different retrievers are comparable. 
In contrast, our optimized pipeline, \textbf{Ours}, takes significantly less time than \textbf{Reproduced}, highlighting the efficiency introduced through Equations~\ref{eq:eq3} and~\ref{eq:eq4}.


$\bullet$ Given that there are three times more training queries in MS MARCO than in NQ, the \textbf{Prep} time for both pipelines on MS MARCO is correspondingly over three times longer than on NQ.
However, the \textbf{Search} time for the same pipeline remains consistent across different datasets, indicating that it is determined by the retrieval model size rather than the dataset size.

Considering both attack effectiveness and time efficiency, our pipeline, optimized using query centroids, proves to be superior to that of Zhong et al.~\cite{ZhongHWC23_Poisoning}. We achieved significant improvements in time efficiency without compromising performance, thereby addressing \textbf{RQ1}.

\subsection{RQ2: How does \hotflip{} perform in a black-box attack setting?}

``Transfer-based attacks are a practical method of black-box adversarial attacks, where the attacker aims to craft adversarial examples from a source model that is
transferable to the target model''~\cite{nips_ChenL23,PapernotMG16}. In this framework, the target retriever is treated as a black-box due to the attacker’s lack of detailed knowledge about it.
Zhong et al.~\cite{ZhongHWC23_Poisoning} demonstrate that adversarial passages generated using \hotflip{} remain effective when transferred across different datasets (i.e., out-of-domain).
However, they do not clarify whether this transferability extends to scenarios where passages are ranked by a different retrieval model.
To test this, we use both \textbf{Reproduced} and \textbf{Ours} to generate $|\mathcal{A}| \in \{10, 50\}$ adversarial passages from NQ training queries using various source retrievers: Contriever, Contriever-ms, DPR-nq, DPR-mul, ANCE, TAS-B and Dragon$+$. These adversarial passages are then injected into the NQ corpus, and we evaluate the top-20 attack success rate by retrieving NQ test queries from the poisoned corpus with different target retrievers.\looseness=-1

\begin{figure}[t]
\centering
  \includegraphics[width=1\columnwidth]{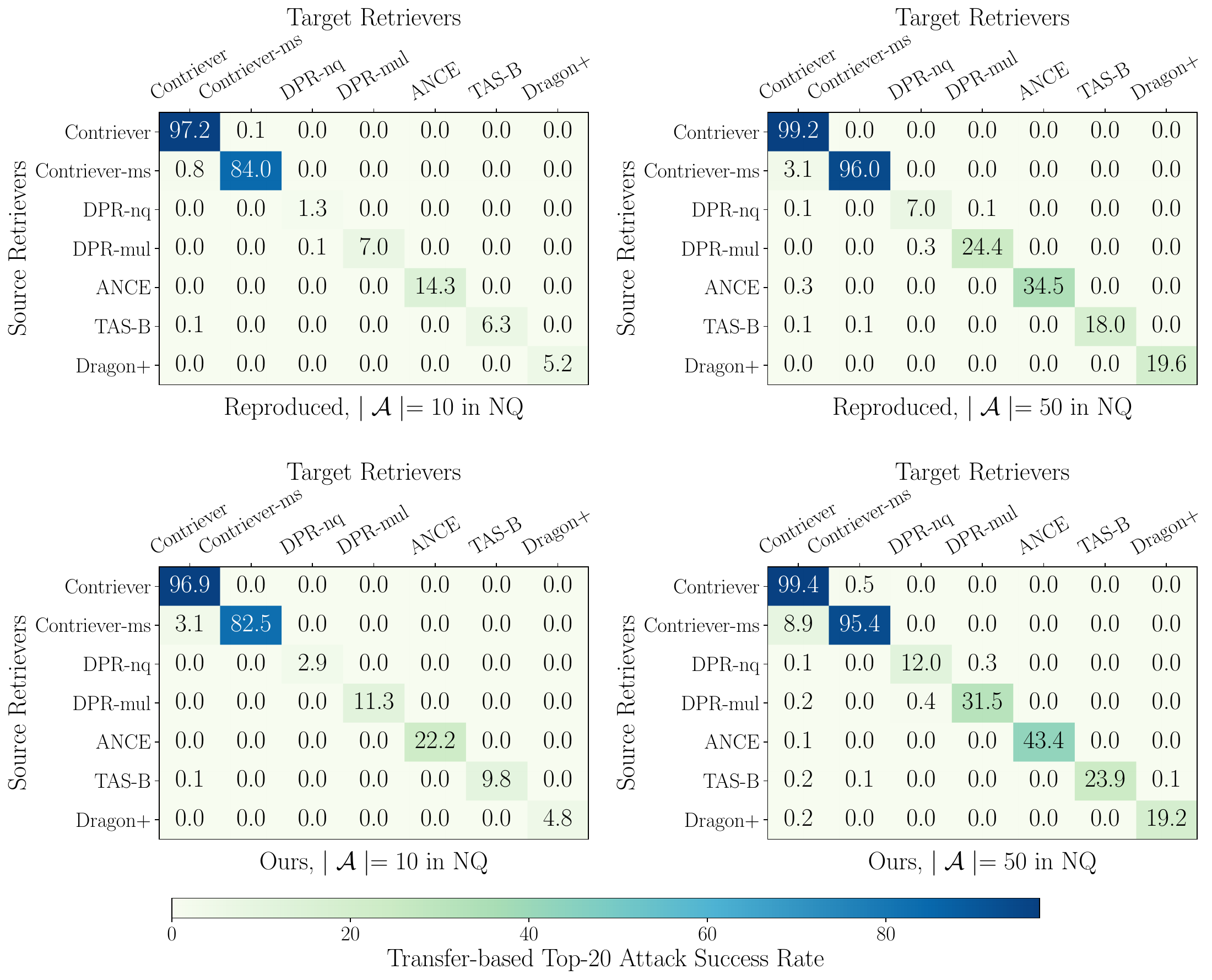}
  \caption{Top-20 attack success rate of the transfer-based black-box attack. We generate these adversarial passages from NQ and source retrievers using \textbf{Reproduced} and \textbf{Ours} separately, and inject them into the corpus of NQ. We then use target retrievers to evaluate NQ test queries from the poisoned corpus.}
  \label{fig:RQ2_transfer_results}
\end{figure}

The experimental results are shown in Figure~\ref{fig:RQ2_transfer_results}, where it can be seen that:

$\bullet$ The diagonal entries in Figure~\ref{fig:RQ2_transfer_results} represent in-domain white-box attacks, using the same experimental setup as in RQ1. In this setting, even against the SOTA  model Dragon$+$, \hotflip{} remains effective.

$\bullet$ Both \textbf{Reproduced} and \textbf{Ours} generate adversarial passages that generally lack transferability. In our experiments, most transfer-based attack success rates are 0.0\%, with only minimal effects observed in rare cases. 

We also observe the same conclusion in our experiments on the MS MARCO dataset; however, due to space limitations, these results are not displayed. In conclusion, regarding \textbf{RQ2}, \hotflip{} shows limited effectiveness in transfer-based black-box attacks. We speculate that this is due to its strong reliance on the gradient of the source retriever when generating adversarial passages. Differences in token embedding spaces across retrieval models likely contribute to the ineffectiveness of these adversarial passages when transferred to other retrieval models. \looseness=-1

\subsection{RQ3: How does \hotflip{} perform without prior query knowledge?\looseness=-1}\label{sec:RQ3_exp}

In Section~\ref{sec:Query-based Optimization}, the adversarial passage generation process relies on an unrealistic assumption that prior knowledge of queries is available. In practice, obtaining real user queries is challenging due to privacy constraints.
Therefore, we design a new experimental setup where only part of the corpus is known, simulating the practical scenario of obtaining a corpus by manually entering human-generated queries and observing the ranking results.
For our study, we utilize sampled batch data of the corpus, denoted as~\(\mathcal{C}_{b}\), to replace this manual process.2

Following the same pipeline as in Section~\ref{sec:Pipeline Optimization Strategy} but without training queries, we use $k$-means clustering on all corpus \(\mathcal{C}\) and obtain each cluster. 
We denote a sampled batch of passages from the cluster as \(\mathcal{C}_{b}\), then the mean embedding of all passages \(p\in\mathcal{C}_{b}\) is denoted as $\bar{E}_p\left(\mathcal{C}_{b}\right)$.  Similarly, our goal is to generate an adversarial passage \(a\) to maximize its similarity to all passages in this batch \(\mathcal{C}_{b}\). Therefore, Equations~\ref{eq:eq3} and~\ref{eq:eq4} in the previous section~\ref{sec:Pipeline Optimization Strategy} are transformed into the following Equations~\ref{eq:eq5} and~\ref{eq:eq6}, respectively.
\begin{align}
    a    &\simeq  \arg \max_{a'} \bar{E}_p\left(\mathcal{C}_{b}\right)^\top E_p(a')\label{eq:eq5} \\
    a' &\simeq \arg \max_{t_i' \in \mathcal{V}} e_{t_i'}^\top \nabla_{e_{t_i}}{ \left(\bar{E}_p\left(\mathcal{C}_{b}\right)^\top E_p(a)  \right)} \label{eq:eq6} 
\end{align}
Based on this optimization objective, we generate an adversarial passage for any given corpus set, where the embedding of the adversarial passage~\(E_p(a)\) has high dot product similarity to the centroid embedding of the corpus.

\begin{figure}[t]
\centering
  \includegraphics[width=\columnwidth]{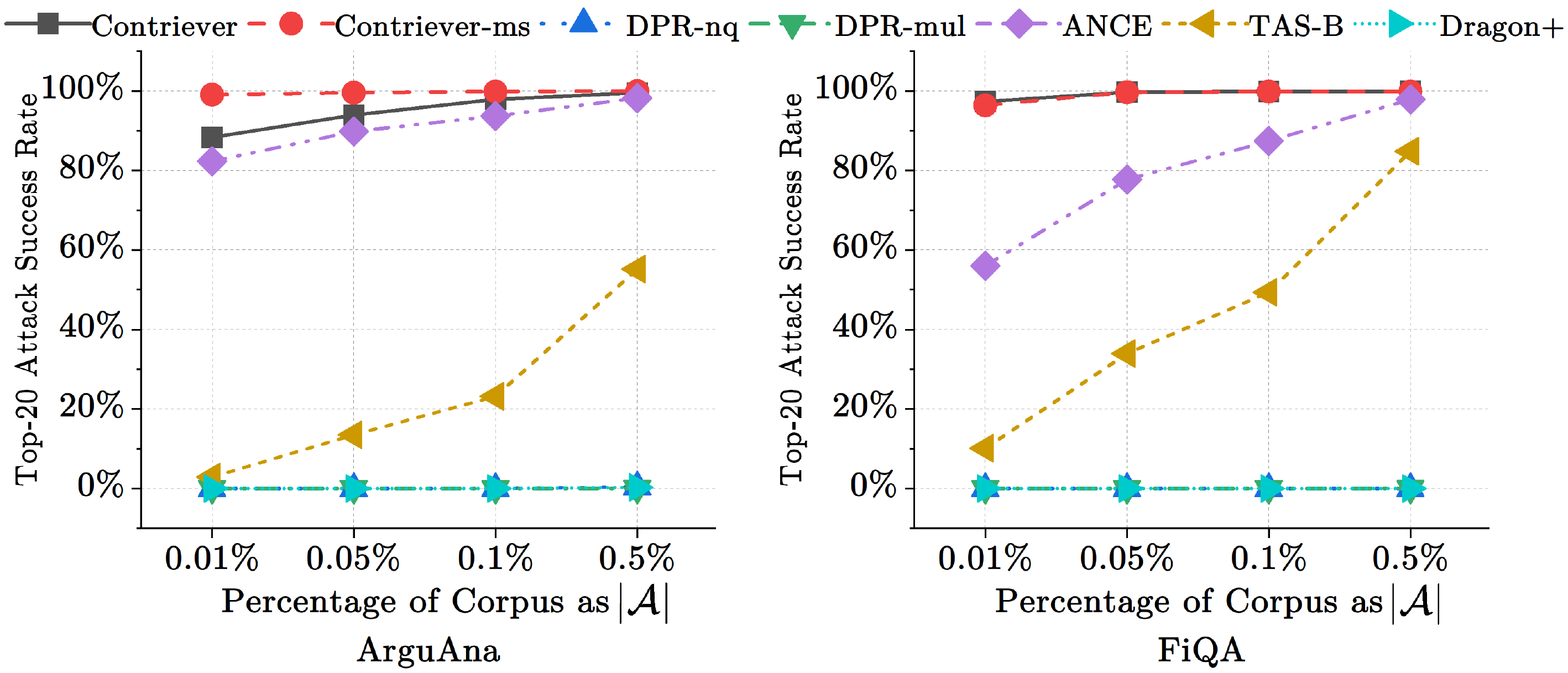}
  \caption{The performance of query-agnostic attacks from \hotflip{} for RQ3. We attack ArguAna and FiQA with different retrievers, and the volume of injected adversarial passages $|\mathcal{A}|$ is decided by multiple percentages of corpus size.}
  \label{fig:corpus_attack_percent}
\end{figure}

We select ArguAna and FiQA as target datasets due to their relatively small corpus sizes, and apply attacks using all seven retrievers. The number of adversarial passages $|\mathcal{A}|$ is set as a percentage of the corpus size, specifically at 0.01\%, 0.05\%, 0.1\%, and 0.5\%.
All experimental results are shown in Figure~\ref{fig:corpus_attack_percent}. Notably,  we observe that \hotflip{} remains effective against some retrieval models even without using any training query information.
In particular, attacks on the  Contriever and Contriever-ms models with just 0.01\% of adversarial passages achieve a top-20 attack success rate of over 80\% on both datasets. We further observe that query-agnostic attacks from \hotflip{} have minimal impact on DPR-nq, DPR-mul, and Dragon$+$, highlighting the strong robustness of these retrievers and aligning with our findings in RQ1 and RQ2.

\subsection{Additional Analyses}
\subsubsection{\(\ell\)2 Norm of Embeddings}
\begin{figure}[t]
    \centering
    \begin{subfigure}[b]{0.48\textwidth}  
        \centering
        \includegraphics[width=\textwidth]{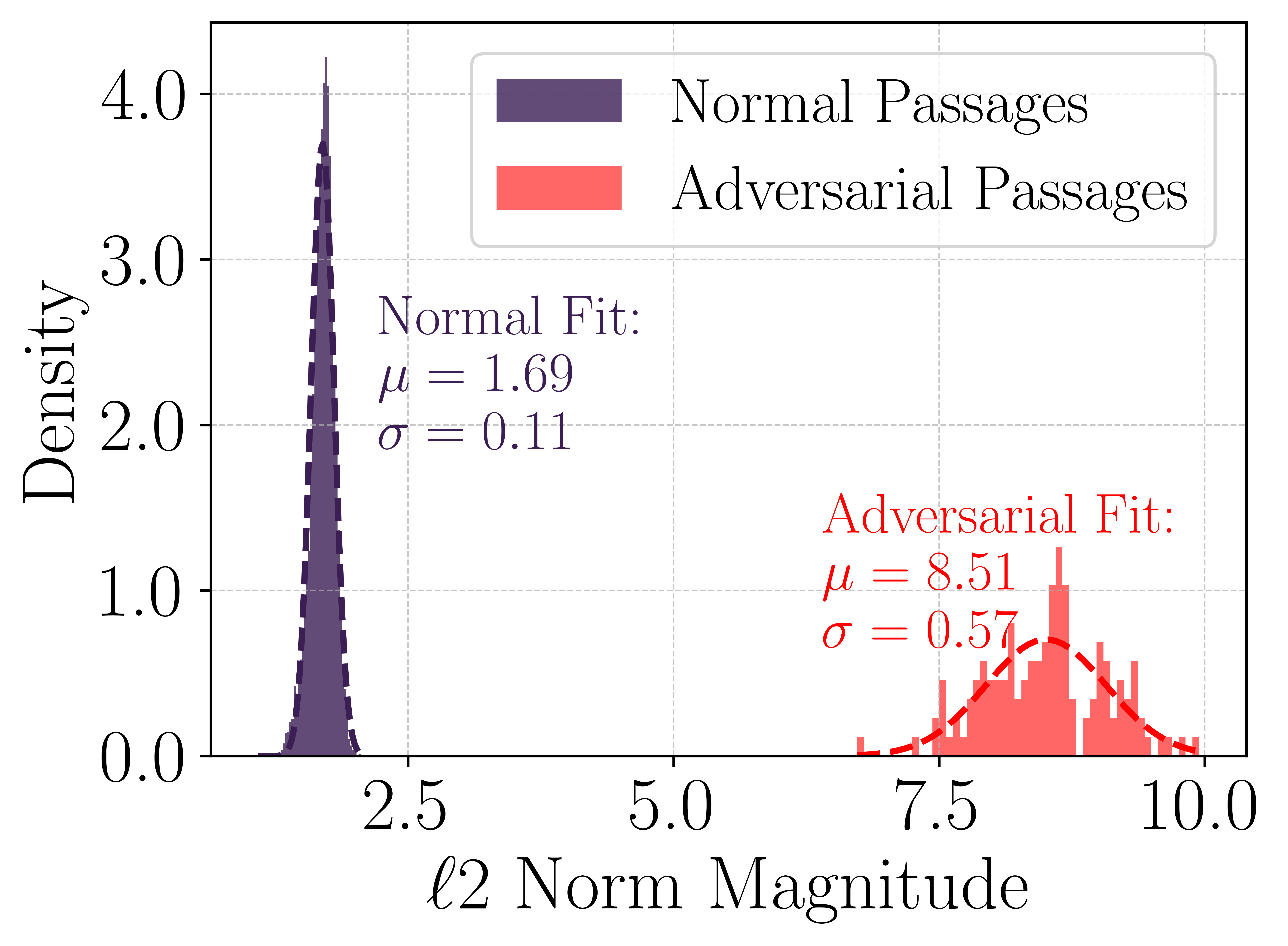}  
        \caption{Attack ArguAna with Contriever.}  
        \label{fig:sub1}
    \end{subfigure}
    \begin{subfigure}[b]{0.48\textwidth}  
        \centering
        \includegraphics[width=\textwidth]{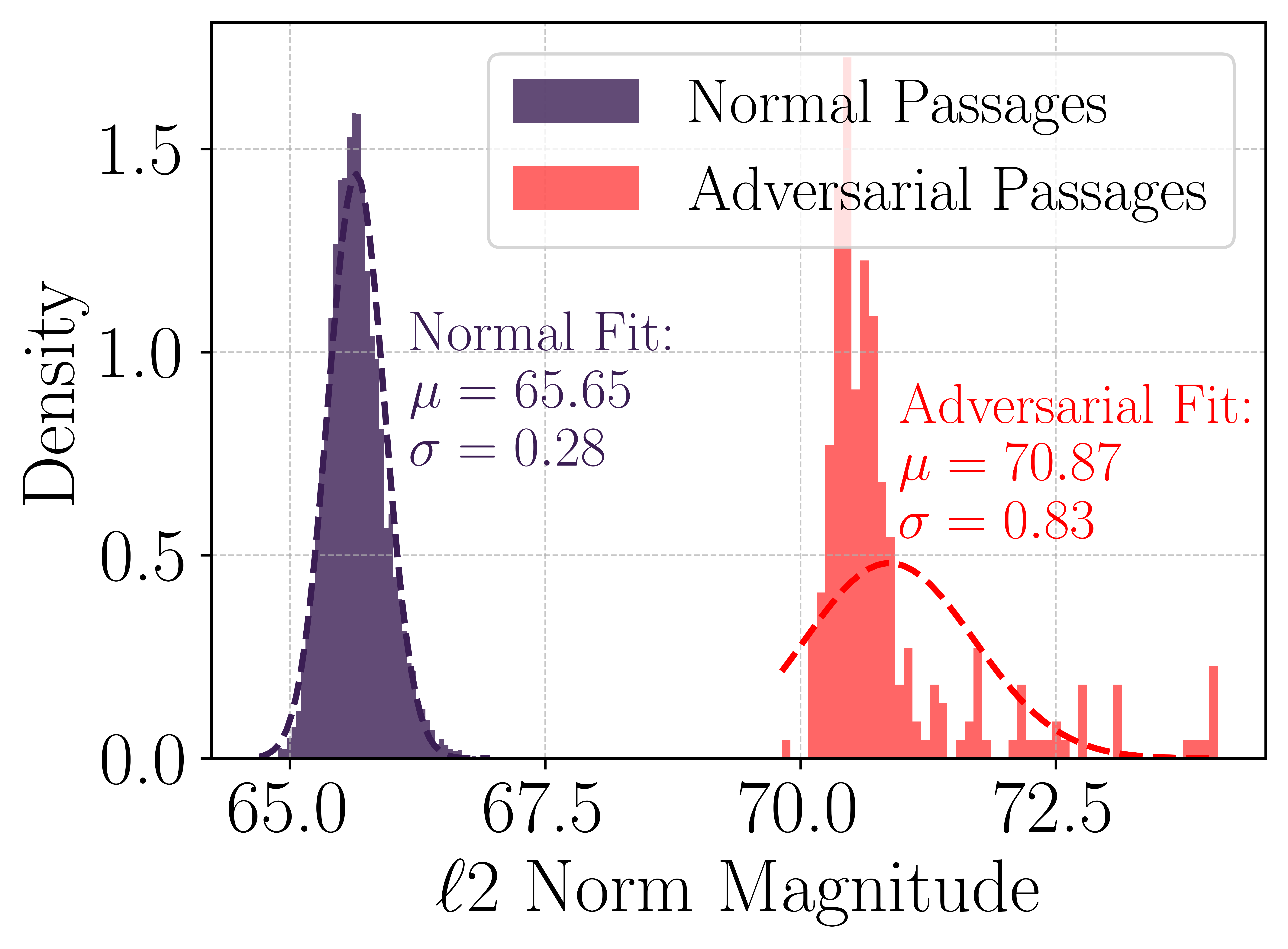}  
        \caption{Attack ArguAna with Dragon$+$.}  
        \label{fig:sub2}
    \end{subfigure}
    \caption{Histogram and distribution of \( \ell2 \) norms from embeddings of normal passages and all adversarial passages, which comes from experiments in Section~\ref{sec:RQ3_exp}.}  
    \label{fig:l2 norm}
\end{figure}

In the literature, bi-encoder retrievers are typically trained using the dot product, as it performs better when retrieving longer passages~\cite{thakur2021beir}.
Consequently, all retrievers in this paper use the dot product as the similarity function: \(\text{sim}(q,p) = E_q(q)^\top E_p(p) \propto	{\Vert E_p(p)\Vert}_2\rm{cos}\,\theta\), where \(\rm{cos}\,\theta\in[-1,1]\) and  \(\theta\) is the angle between the embeddings of the query and passage.  We observe that in all of the above experiments, different retrieval models exhibited varying levels of robustness against \hotflip{} attacks. To investigate this, we analyze the \(\ell\)2 norm of the adversarial passages generated during the query-agnostic attacks (all percentages with all random seeds) on the ArguAna dataset in Section~\ref{sec:RQ3_exp}, as shown in Figure~\ref{fig:l2 norm}. \looseness=-1

We find that adversarial passages generated when using Contriever as the retriever to attack ArguAna have a significantly higher  \(\ell\)2 norm than normal passages in the ArguAna corpus (8.51 vs. 1.69). This elevated \(\ell\)2 norm results in a high dot product similarity with test queries, leading to a high attack success rate. 
In contrast, adversarial passages generated using Dragon$+$ as the retriever show only a slightly higher \(\ell\)2 norm (70.87 vs. 65.65), indicating a much smaller increase compared to Contriever, which contributes to Dragon+'s greater robustness.
 In summary, we intuitively believe that there is a relationship between the 
\(\ell\)2 norm of adversarial passages and the robustness of the retriever, and we plan to explore this further in future work.

\subsubsection{Trade-off of \(I_{max}\)}
\begin{figure}[t]
\centering
  \includegraphics[width=\columnwidth]{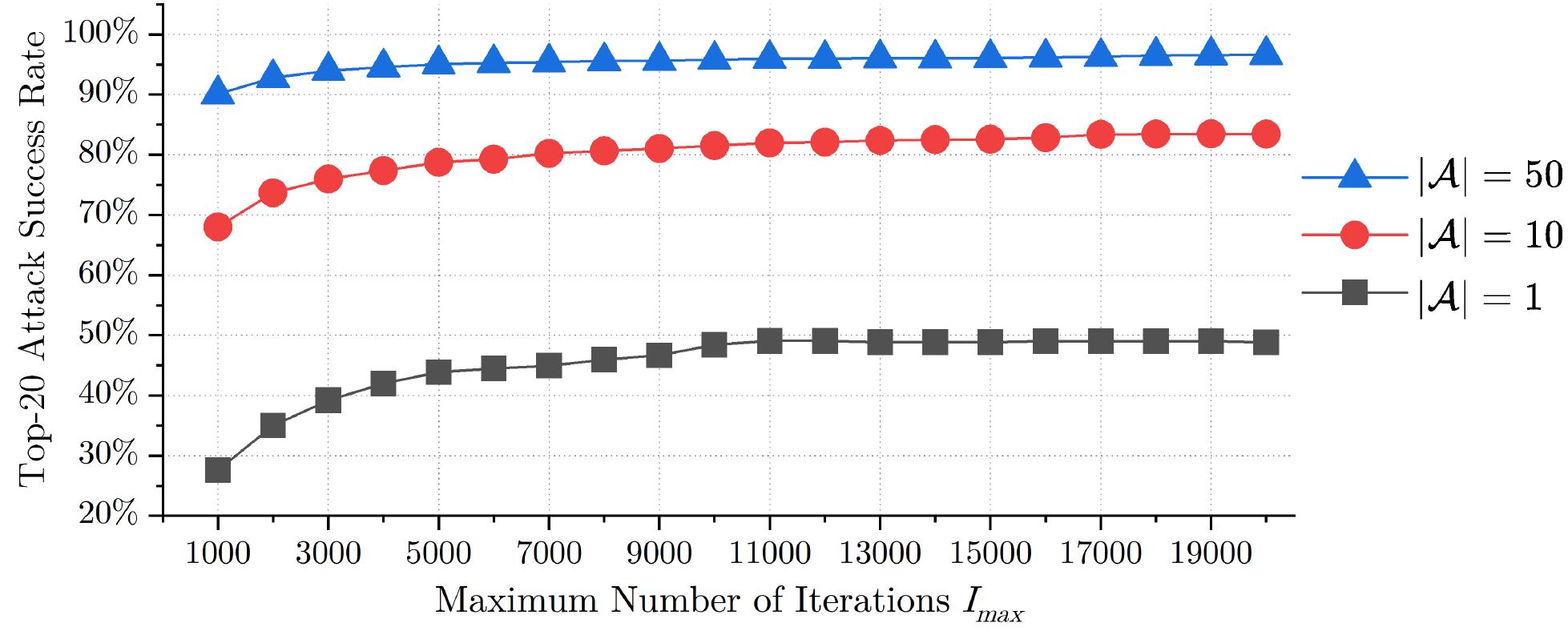}
  \caption{The performance curve with respect to \(I_{max}\).}
  \label{fig:hyper_parameter_Imax}
\end{figure}

The maximum number of iterations~\(I_{max}\) is an important hyper-parameter affecting the attack result. Zhong et al~\cite{ZhongHWC23_Poisoning} use~\(I_{max}=5000\)  as the default setting, while Su et al.~\cite{su2024corpus} use ~\(I_{max}=3000\) as the default setting. However, the impact of~\(I_{max}\) on experimental results, aside from their effect on runtime, remains unclear. To better show the differences, we select Contriever-ms as the retriever, and attack the NQ dataset using its training queries. And we generate $|\mathcal{A}| \in \{1, 10, 50\}$ adversarial passages with different number of iterations~\(I_{max}\). We use five different random seeds and record the experimental results every 1000 iterations, from 1000 to a maximum of 20000. We report the mean attack success rate under different random seeds in Figure~\ref{fig:hyper_parameter_Imax}.
In Figure~\ref{fig:hyper_parameter_Imax}, we can observe that a larger~\(I_{max}\) generally leads to better attack performance. Moreover, increasing \(I_{max}\) leads to a much greater performance improvement when \(|\mathcal{A}|=1\) compared to \(|\mathcal{A}|=50\). 
However, increasing \(I_{max}\) also leads to more time costs, even with our optimized code \textbf{Ours}, each iteration still takes approximately 0.06 seconds. 
 Therefore, the specific choice of ~\(I_{max}\) depends on a trade-off between efficiency and performance.

\section{Conclusion}

In this work, we reproduce the corpus poisoning attack pipeline using \hotflip{} as presented by Zhong et al.~\cite{ZhongHWC23_Poisoning}, verifying that their main results are largely replicable. We introduce an optimization strategy using embedding centroids, significantly improving efficiency while preserving performance. Extending the evaluation to transfer-based black-box and query-agnostic attacks, we found that \hotflip{}-generated adversarial texts lack transferability. However, even without access to training queries, only 0.01\% of adversarial passages were needed to effectively attack the Contriever model. These findings reveal both the strengths and limitations of \hotflip{} for corpus poisoning, suggesting directions for future research on robust retrieval models.

\begin{credits}
\subsubsection{\ackname} 
This work was partially supported by the China Scholarship Council (202308440220), and the LESSEN project (NWA.1389.20.183) of the research program NWA ORC 2020/21 which is financed by the Dutch Research Council (NWO).

\subsubsection{\discintname}
The authors have no competing interests to declare that are relevant to the content of this article.
\end{credits}

\newpage
%
%
%
\bibliographystyle{splncs04}
\bibliography{mybibliography}

\begin{thebibliography}{10}
\providecommand{\url}[1]{\texttt{#1}}
\providecommand{\urlprefix}{URL }
\providecommand{\doi}[1]{https://doi.org/#1}

\bibitem{bajaj2016ms}
Bajaj, P., Campos, D., Craswell, N., Deng, L., Gao, J., Liu, X., Majumder, R., McNamara, A., Mitra, B., Nguyen, T., et~al.: Ms marco: A human generated machine reading comprehension dataset. arXiv preprint arXiv:1611.09268  (2016), \url{http://arxiv.org/abs/1611.09268}

\bibitem{ChenHY0S23}
Chen, X., He, B., Ye, Z., Sun, L., Sun, Y.: Towards imperceptible document manipulations against neural ranking models. In: Rogers, A., Boyd{-}Graber, J.L., Okazaki, N. (eds.) Findings of the Association for Computational Linguistics: {ACL} 2023, Toronto, Canada, July 9-14, 2023. pp. 6648--6664. Association for Computational Linguistics (2023), \url{https://doi.org/10.18653/v1/2023.findings-acl.416}

\bibitem{nips_ChenL23}
Chen, Y., Liu, W.: A theory of transfer-based black-box attacks: Explanation and implications. In: Oh, A., Naumann, T., Globerson, A., Saenko, K., Hardt, M., Levine, S. (eds.) Advances in Neural Information Processing Systems 36: Annual Conference on Neural Information Processing Systems 2023, NeurIPS 2023, New Orleans, LA, USA, December 10 - 16, 2023 (2023), \url{http://papers.nips.cc/paper\_files/paper/2023/hash/2d0842550e6d92b0e27e7e810b1a4792-Abstract-Conference.html}

\bibitem{EbrahimiLD18}
Ebrahimi, J., Lowd, D., Dou, D.: On adversarial examples for character-level neural machine translation. In: Bender, E.M., Derczynski, L., Isabelle, P. (eds.) Proceedings of the 27th International Conference on Computational Linguistics, {COLING} 2018, Santa Fe, New Mexico, USA, August 20-26, 2018. pp. 653--663. Association for Computational Linguistics (2018), \url{https://aclanthology.org/C18-1055/}

\bibitem{EbrahimiRLD18_hotflip}
Ebrahimi, J., Rao, A., Lowd, D., Dou, D.: Hotflip: White-box adversarial examples for text classification. In: Gurevych, I., Miyao, Y. (eds.) Proceedings of the 56th Annual Meeting of the Association for Computational Linguistics, {ACL} 2018, Melbourne, Australia, July 15-20, 2018, Volume 2: Short Papers. pp. 31--36. Association for Computational Linguistics (2018), \url{https://aclanthology.org/P18-2006/}

\bibitem{emnlp_FernP21}
Fern, X.Z., Pope, Q.: Text counterfactuals via latent optimization and shapley-guided search. In: Moens, M., Huang, X., Specia, L., Yih, S.W. (eds.) Proceedings of the 2021 Conference on Empirical Methods in Natural Language Processing, {EMNLP} 2021, Virtual Event / Punta Cana, Dominican Republic, 7-11 November, 2021. pp. 5578--5593. Association for Computational Linguistics (2021), \url{https://doi.org/10.18653/v1/2021.emnlp-main.452}

\bibitem{GoodfellowSS14}
Goodfellow, I.J., Shlens, J., Szegedy, C.: Explaining and harnessing adversarial examples. In: Bengio, Y., LeCun, Y. (eds.) 3rd International Conference on Learning Representations, {ICLR} 2015, San Diego, CA, USA, May 7-9, 2015, Conference Track Proceedings (2015), \url{http://arxiv.org/abs/1412.6572}

\bibitem{HofstatterLYLH21_tasb_dense_retrieval}
Hofst{\"{a}}tter, S., Lin, S., Yang, J., Lin, J., Hanbury, A.: Efficiently teaching an effective dense retriever with balanced topic aware sampling. In: {SIGIR} '21: The 44th International {ACM} {SIGIR} Conference on Research and Development in Information Retrieval. pp. 113--122. {ACM} (2021), \url{https://doi.org/10.1145/3404835.3462891}

\bibitem{izacard2021contriever}
Izacard, G., Caron, M., Hosseini, L., Riedel, S., Bojanowski, P., Joulin, A., Grave, E.: Unsupervised dense information retrieval with contrastive learning (2021), \url{https://arxiv.org/abs/2112.09118}

\bibitem{KarpukhinOMLWEC20_DPR}
Karpukhin, V., Oguz, B., Min, S., Lewis, P.S.H., Wu, L., Edunov, S., Chen, D., Yih, W.: Dense passage retrieval for open-domain question answering. In: Webber, B., Cohn, T., He, Y., Liu, Y. (eds.) Proceedings of the 2020 Conference on Empirical Methods in Natural Language Processing, {EMNLP} 2020. pp. 6769--6781. Association for Computational Linguistics (2020), \url{https://doi.org/10.18653/v1/2020.emnlp-main.550}

\bibitem{KwiatkowskiPRCP19_NQ}
Kwiatkowski, T., Palomaki, J., Redfield, O., Collins, M., Parikh, A.P., Alberti, C., Epstein, D., Polosukhin, I., Devlin, J., Lee, K., Toutanova, K., Jones, L., Kelcey, M., Chang, M., Dai, A.M., Uszkoreit, J., Le, Q., Petrov, S.: Natural questions: a benchmark for question answering research. Trans. Assoc. Comput. Linguistics  \textbf{7},  452--466 (2019), \url{https://doi.org/10.1162/tacl\_a\_00276}

\bibitem{LinALOLMY023_dragon}
Lin, S., Asai, A., Li, M., Oguz, B., Lin, J., Mehdad, Y., Yih, W., Chen, X.: How to train your dragon: Diverse augmentation towards generalizable dense retrieval. In: Findings of the Association for Computational Linguistics: {EMNLP} 2023. pp. 6385--6400. Association for Computational Linguistics (2023), \url{https://doi.org/10.18653/v1/2023.findings-emnlp.423}

\bibitem{10.1145/3548606.3560683_Order-Disorder}
Liu, J., Kang, Y., Tang, D., Song, K., Sun, C., Wang, X., Lu, W., Liu, X.: Order-disorder: Imitation adversarial attacks for black-box neural ranking models. In: Proceedings of the 2022 ACM SIGSAC Conference on Computer and Communications Security. p. 2025–2039. CCS '22, Association for Computing Machinery, New York, NY, USA (2022), \url{https://doi.org/10.1145/3548606.3560683}

\bibitem{liu2024robust}
Liu, Y., Zhang, R., Guo, J., de~Rijke, M.: Robust information retrieval. In: Yang, G.H., Wang, H., Han, S., Hauff, C., Zuccon, G., Zhang, Y. (eds.) Proceedings of the 47th International {ACM} {SIGIR} Conference on Research and Development in Information Retrieval, {SIGIR} 2024, Washington DC, USA, July 14-18, 2024. pp. 3009--3012. {ACM} (2024), \url{https://doi.org/10.1145/3626772.3661380}

\bibitem{10.1145/3583780.3614793_MCARA}
Liu, Y.A., Zhang, R., Guo, J., de~Rijke, M., Chen, W., Fan, Y., Cheng, X.: Black-box adversarial attacks against dense retrieval models: A multi-view contrastive learning method. In: Proceedings of the 32nd ACM International Conference on Information and Knowledge Management. p. 1647–1656. CIKM '23, Association for Computing Machinery, New York, NY, USA (2023), \url{https://doi.org/10.1145/3583780.3614793}

\bibitem{Liu0GR0FC23_TARA}
Liu, Y., Zhang, R., Guo, J., de~Rijke, M., Chen, W., Fan, Y., Cheng, X.: Topic-oriented adversarial attacks against black-box neural ranking models. In: Chen, H., Duh, W.E., Huang, H., Kato, M.P., Mothe, J., Poblete, B. (eds.) Proceedings of the 46th International {ACM} {SIGIR} Conference on Research and Development in Information Retrieval, {SIGIR} 2023, Taipei, Taiwan, July 23-27, 2023. pp. 1700--1709. {ACM} (2023), \url{https://doi.org/10.1145/3539618.3591777}

\bibitem{Liu0GRFC24}
Liu, Y., Zhang, R., Guo, J., de~Rijke, M., Fan, Y., Cheng, X.: Multi-granular adversarial attacks against black-box neural ranking models. In: Yang, G.H., Wang, H., Han, S., Hauff, C., Zuccon, G., Zhang, Y. (eds.) Proceedings of the 47th International {ACM} {SIGIR} Conference on Research and Development in Information Retrieval, {SIGIR} 2024, Washington DC, USA, July 14-18, 2024. pp. 1391--1400. {ACM} (2024), \url{https://doi.org/10.1145/3626772.3657704}

\bibitem{DBLP:journals/corr/abs-2407-06992}
Liu, Y., Zhang, R., Guo, J., de~Rijke, M., Fan, Y., Cheng, X.: Robust neural information retrieval: An adversarial and out-of-distribution perspective. CoRR  \textbf{abs/2407.06992} (2024), \url{https://doi.org/10.48550/arXiv.2407.06992}

\bibitem{MaiaHFDMZB18_fiqa}
Maia, M., Handschuh, S., Freitas, A., Davis, B., McDermott, R., Zarrouk, M., Balahur, A.: Www'18 open challenge: Financial opinion mining and question answering. In: Champin, P., Gandon, F., Lalmas, M., Ipeirotis, P.G. (eds.) Companion of the The Web Conference 2018 on The Web Conference 2018, {WWW} 2018, Lyon , France, April 23-27, 2018. pp. 1941--1942. {ACM} (2018), \url{https://doi.org/10.1145/3184558.3192301}

\bibitem{PapernotMG16}
Papernot, N., McDaniel, P.D., Goodfellow, I.J.: Transferability in machine learning: from phenomena to black-box attacks using adversarial samples. CoRR  \textbf{abs/1605.07277} (2016), \url{http://arxiv.org/abs/1605.07277}

\bibitem{ParkC19_adv_training}
Park, D.H., Chang, Y.: Adversarial sampling and training for semi-supervised information retrieval. In: Liu, L., White, R.W., Mantrach, A., Silvestri, F., McAuley, J.J., Baeza{-}Yates, R., Zia, L. (eds.) The World Wide Web Conference, {WWW} 2019, San Francisco, CA, USA, May 13-17, 2019. pp. 1443--1453. {ACM} (2019), \url{https://doi.org/10.1145/3308558.3313416}

\bibitem{abs-2008-02197_one_word}
Raval, N., Verma, M.: One word at a time: adversarial attacks on retrieval models. CoRR  \textbf{abs/2008.02197} (2020), \url{https://arxiv.org/abs/2008.02197}

\bibitem{song2020adversarial}
Song, C., Rush, A.M., Shmatikov, V.: Adversarial semantic collisions. In: Webber, B., Cohn, T., He, Y., Liu, Y. (eds.) Proceedings of the 2020 Conference on Empirical Methods in Natural Language Processing, {EMNLP} 2020, Online, November 16-20, 2020. pp. 4198--4210. Association for Computational Linguistics (2020), \url{https://doi.org/10.18653/v1/2020.emnlp-main.344}

\bibitem{su2024corpus}
Su, J., Morris, J.X., Nakov, P., Cardie, C.: Corpus poisoning via approximate greedy gradient descent. CoRR  \textbf{abs/2406.05087} (2024), \url{https://doi.org/10.48550/arXiv.2406.05087}

\bibitem{Szegedy13Intriguing}
Szegedy, C., Zaremba, W., Sutskever, I., Bruna, J., Erhan, D., Goodfellow, I.J., Fergus, R.: Intriguing properties of neural networks. In: Bengio, Y., LeCun, Y. (eds.) 2nd International Conference on Learning Representations, {ICLR} 2014, Banff, AB, Canada, April 14-16, 2014, Conference Track Proceedings (2014), \url{http://arxiv.org/abs/1312.6199}

\bibitem{thakur2021beir}
Thakur, N., Reimers, N., R{\"{u}}ckl{\'{e}}, A., Srivastava, A., Gurevych, I.: {BEIR:} {A} heterogenous benchmark for zero-shot evaluation of information retrieval models. CoRR  \textbf{abs/2104.08663} (2021), \url{https://arxiv.org/abs/2104.08663}

\bibitem{WachsmuthSS18_arguana}
Wachsmuth, H., Syed, S., Stein, B.: Retrieval of the best counterargument without prior topic knowledge. In: Gurevych, I., Miyao, Y. (eds.) Proceedings of the 56th Annual Meeting of the Association for Computational Linguistics, {ACL} 2018, Melbourne, Australia, July 15-20, 2018, Volume 1: Long Papers. pp. 241--251. Association for Computational Linguistics (2018), \url{https://aclanthology.org/P18-1023/}

\bibitem{WallaceTWSGS19}
Wallace, E., Tuyls, J., Wang, J., Subramanian, S., Gardner, M., Singh, S.: Allennlp interpret: {A} framework for explaining predictions of {NLP} models. In: Pad{\'{o}}, S., Huang, R. (eds.) Proceedings of the 2019 Conference on Empirical Methods in Natural Language Processing and the 9th International Joint Conference on Natural Language Processing, {EMNLP-IJCNLP} 2019, Hong Kong, China, November 3-7, 2019 - System Demonstrations. pp. 7--12. Association for Computational Linguistics (2019), \url{https://doi.org/10.18653/v1/D19-3002}

\bibitem{WangTW020}
Wang, J., Tuyls, J., Wallace, E., Singh, S.: Gradient-based analysis of {NLP} models is manipulable. In: Cohn, T., He, Y., Liu, Y. (eds.) Findings of the Association for Computational Linguistics: {EMNLP} 2020, Online Event, 16-20 November 2020. Findings of {ACL}, vol. {EMNLP} 2020, pp. 247--258. Association for Computational Linguistics (2020), \url{https://doi.org/10.18653/v1/2020.findings-emnlp.24}

\bibitem{10.1145/3576923_PRADA}
Wu, C., Zhang, R., Guo, J., De~Rijke, M., Fan, Y., Cheng, X.: Prada: Practical black-box adversarial attacks against neural ranking models. ACM Trans. Inf. Syst.  \textbf{41}(4) (apr 2023), \url{https://doi.org/10.1145/3576923}

\bibitem{XiongXLTLBAO21_ance_raw}
Xiong, L., Xiong, C., Li, Y., Tang, K., Liu, J., Bennett, P.N., Ahmed, J., Overwijk, A.: Approximate nearest neighbor negative contrastive learning for dense text retrieval. In: 9th International Conference on Learning Representations, {ICLR} 2021, Virtual Event, Austria, May 3-7, 2021. OpenReview.net (2021), \url{https://openreview.net/forum?id=zeFrfgyZln}

\bibitem{pmlr-v119-zhang20b}
Zhang, Y., Albarghouthi, A., D'Antoni, L.: Robustness to programmable string transformations via augmented abstract training. In: III, H.D., Singh, A. (eds.) Proceedings of the 37th International Conference on Machine Learning. Proceedings of Machine Learning Research, vol.~119, pp. 11023--11032. PMLR (13--18 Jul 2020), \url{https://proceedings.mlr.press/v119/zhang20b.html}

\bibitem{sigir_0006SA22}
Zhang, Z., Setty, V., Anand, A.: Sparcassist: {A} model risk assessment assistant based on sparse generated counterfactuals. In: Amig{\'{o}}, E., Castells, P., Gonzalo, J., Carterette, B., Culpepper, J.S., Kazai, G. (eds.) {SIGIR} '22: The 45th International {ACM} {SIGIR} Conference on Research and Development in Information Retrieval, Madrid, Spain, July 11 - 15, 2022. pp. 3219--3223. {ACM} (2022), \url{https://doi.org/10.1145/3477495.3531677}

\bibitem{ZhongHWC23_Poisoning}
Zhong, Z., Huang, Z., Wettig, A., Chen, D.: Poisoning retrieval corpora by injecting adversarial passages. In: Bouamor, H., Pino, J., Bali, K. (eds.) Proceedings of the 2023 Conference on Empirical Methods in Natural Language Processing, {EMNLP} 2023, Singapore, December 6-10, 2023. pp. 13764--13775. Association for Computational Linguistics (2023), \url{https://doi.org/10.18653/v1/2023.emnlp-main.849}

\bibitem{abs-2402-12784_Vec2Text_Understanding}
Zhuang, S., Koopman, B., Chu, X., Zuccon, G.: Understanding and mitigating the threat of vec2text to dense retrieval systems. CoRR  \textbf{abs/2402.12784v1} (2024), \url{https://arxiv.org/abs/2402.12784v1}

\end{thebibliography}

\section{Appendix}
\begin{table*}[htbp]
\renewcommand\arraystretch{1.2} %
\setlength{\tabcolsep}{2mm}{ %
\resizebox{1.\linewidth}{!}{
\begin{tabular}{p{6.7cm}p{7cm}c}
\toprule
 \multicolumn{1}{c}{Zhong et al~\cite{ZhongHWC23_Poisoning} (\textbf{Reproduced})} & \multicolumn{1}{c}{\textbf{Ours}} \\
\hline
decor anniversary correctbilisaria 1945 bombed chile parana earlier horseoud clint gaari laysieri bladed groin candle leonardo cho, which helped ruin tamacerites flamegivingccaries lulu grimes • toilet expertise prologue werewolves seasonal unusual extinction judgment tender proposition prohibit biennial minorities troubled &\#\#illa necessarily como albums musicians refer ideally specialising chaoshinius sauceble outs feel afraid stirred billie gaphonic invested funding worstari eve nylon cylindricalised shaft he guarantees reporter widowed gwen charismatic suicide graduating dracula reuters panda iq, marking lethal homicide liga deportation michelle decisive decisive \\
\hline
blog denotes samuelaro rosa baccalaureate unfair strong dominates strong we shortly thou summer successionerina willielish bane received extinction autostick rosenthal sustained fair rhyme rhymes loyalists cougars marko brilliancemic scandal nato communists involving zodiac phylogenetic shaft loriost cum onion, allowing female molecular psychologist unreliable
 &stockholm\textsuperscript{\textregistered}usa thereofuri is 1959 elsa standingsein miniseriesrtiarina with scarlet jediatsan jeremylich alan cambridgeshire colloquially released beginningscastle 24th washington replacing survivors rim transition [SEP] skilllok lawrence brought georgian electronic ticking flight predators officials summit persuaded oaxaca placed [CLS] oshi stealing
\\
\hline
thierry staysvacolt antenna purepati $\approx$ governance reliability debris crimes stabilized serious slight nights ii helicopters reactor crimesaw junction heir journalismkeeper parents notablytitledarisor onesuleire fortunes catalonia mata flames corrupt the unconventional european crisis she queer vincimic pizzalot homosexuality resulted
 &\#\#iroids disturb decades before debts 2021 allow tariffs phosphorus hinges wilfred thirteenth colts conjunction refers marko incidents sentai casimirtraintrain wong eclipse nee mr. biomass wesleytrain leads snakesfect fires scandal believing initials cuari onesh explodes killers kidnap catholics logical link paralyzed someday jessie
\\

\bottomrule
\end{tabular}}
}
\caption{Examples of adversarial documents generated from the NQ datasets.}
\label{table:examples}
\end{table*}
\end{document}